\documentclass[10pt]{article}

\usepackage{amsmath}
\usepackage{amssymb}
\usepackage{graphicx}
\usepackage{cite}
\usepackage{color} 

\usepackage[labelfont=bf,labelsep=period,justification=raggedright]{caption}

\makeatletter
\renewcommand{\@biblabel}[1]{\quad#1.}
\makeatother

\date{}

\begin{document}

\begin{flushleft}
{\Large
\textbf{Dworkin's Paradox}
}
\\
Seung Ki Baek$^{1}$,
Jung-Kyoo Choi$^{2}$, 
Beom Jun Kim$^{3,\ast}$
\\
\bf{1} Integrated Science Laboratory, Ume{\aa} University, S-901 87
Ume{\aa}, Sweden (Present address: School of Physics, Korea Institute for
Advanced Study, Seoul 130-722, Korea)
\\
\bf{2} School of Economics and Trade, Kyungpook National University, Daegu
702-701, Korea
\\
\bf{3} BK21 Physics Research Division and Department of Physics,
Sungkyunkwan University, Suwon 440-746, Korea
\\
$\ast$ E-mail: beomjun@skku.edu
\end{flushleft}

\section*{Abstract}
How to distribute welfare in a society is a key issue in the subject of
distributional justice, which is deeply involved with notions of
fairness. Following a thought experiment by Dworkin,
this work considers a society of individuals with different
preferences on the welfare distribution and an official to mediate the
coordination among them. Based on a simple assumption that an individual's
welfare is proportional to how her preference is fulfilled by the actual
distribution, we show that an egalitarian preference is a strict Nash
equilibrium and can be favorable even in certain inhomogeneous situations.
These suggest how communication can encourage and secure a notion of
fairness.

\section*{Introduction}

\subsection*{Background}

The concept of distributive justice has been extensively studied in
political philosophy and economics over the past few decades. One of the
most important milestones in this field is Rawls' {\it A Theory of
Justice}~\cite{rawls}, which
put forward equality as an outcome of the social contract that individuals
behind a veil of ignorance should agree on. Even though egalitarian
doctrine that all human persons are equal in fundamental worth or moral
status is a commonly shared idea, egalitarianism turns out to be a contested
concept. There have been several divergent understandings of the meaning of
equality, ways to achieve equality, or the metric to measure
equality~\cite{sen,sen2}.
For instance, someone who puts more emphasis on equality of
opportunity may have a very different opinion from those who put emphasis on
equality of incomes in spite of the overall agreement on the concept of
equality per se.

Among numerous dimensions where egalitarianism varies, the question of what
should be equalized is one
that many different theories are competing on; is it opportunity,
capabilities, resource or welfare~\cite{sen,sen2,dw1,dw2,hausman}?
Dworkin, one of the most influential proponents of resource egalitarianism,
admits the immediate appeal of the idea that it must ultimately be equality
of welfare insofar as equality is important, and examines the logical
consistency and practical applicability of this welfare
egalitarianism~\cite{dw}.
According to Dworkin, welfare egalitarianism, concerned
with equality in every person's overall satisfaction, has an inconsistency
in its logic. For example, if one accepts the idea that those who are
handicapped need more resources to achieve equal welfare, the same argument
should apply to those who have expensive tastes for the same reason.
However, one should immediately recognize that the appeal of welfare
egalitarianism becomes much less strong in the case of expensive tastes
than in the case of the handicapped. The fact that the same idea can be
accepted in some cases and seems disturbing in other cases reveals a logical
inconsistency of welfare egalitarianism. Dworkin also criticized
welfare-based egalitarianism on that it inevitably relies on the possibility
of interpersonal comparisons of utility, which places a large burden to a
policy maker in practice. Lastly, Dworkin argues that it would probably
prove impossible to reach a reasonable degree of equality in this conception
in a community whose members held very different and very deeply felt
political theories about justice in distribution. 

The last point is the one that we focus on in this paper. We will show that
the existence of contradictory political theories does not immediately
lead to the impossibility but can be formulated as dynamics which admits a
unique solution under certain assumptions.
By doing this, we will argue that the reasoning in \cite{dw} can be
regarded as a tool to analyze and advocate the idea of equality in welfare.
To some extent, this is complementary to a previous work which argues that
one can reach the idea of equality in welfare by starting from that of
equality in resources~\cite{roemer}.
Following the logic that Dworkin used when he showed the
impossibility to reach an agreement on redistribution in terms of welfare,
we also set aside the issues of logical inconsistency and inter-personal
comparisons of individuals' welfare. To focus on the relationship
between individual preferences and resulting welfare, we also set aside the
issue of impartiality (see, e.g., \cite{clip,mongin}).
We will further assume that
individuals have preferences over the distribution of welfare among
them~\cite{fehr,bolton}.
Many theoretical and
experimental studies have shown that people are concerned with equality and
fairness and often persist in fairness even when they lose monetary payoffs
in doing so, e.g., in the public games, ultimatum games or dictator games.
This behavior cannot be explained based on the assumption of
self-regarding preferences but of others-regarding preferences or social
preferences. Our formulation in this paper can serve as a systematic
description for such an approach, because it tells us a way to translate
others' payoffs into an individual's with respect to individual preference.

\subsection*{Model}

Let the concept of welfare be understood as the fulfillment of
preferences~\cite{singer}, including success in political
preferences, i.e., opinions of how welfare should be distributed. Then the
reasoning by Dworkin~\cite{dw}
argues that such an egalitarian society, where everyone is
concerned with the equality, will end up with supporting non-egalitarians by
its own logic. Suppose that a bigot enters an egalitarian society, with an
opinion that some people deserve more than the others. This person will feel
frustrated to see that her political preferences are not accepted by
egalitarian neighbors, and her welfare becomes relatively lower than the
others'. If there is an official committed to compensating for inequality in
welfare, by reallocating resources for example, the bigot should get extra
resources from the official due to her political frustration, {\em because}
she does not support the egalitarian idea of the society. This is called
Dworkin's paradox in this work. Particularly we note that it can serve as an
idealized model to represent our understanding of a modern democratic
society. We will look into this hypothetical society a little
closer.

Imagine a society of $N$ persons and an official. The official,
representing a social institution, exists to
mediate the global coordination. We assume
that the total amount of welfare to be distributed among the persons is
fixed as unity, and that the welfare is infinitely divisible, since we are
interested only in relative fractions rather than absolute amounts that
individuals have. The official herself does not take part in sharing the
welfare, but only receives the $N$ persons' opinions and find a way to
distribute the welfare among them.
Let each person $i$ have a certain preference about how the welfare should
be distributed, say $\mathbf{v}_i = (v_{i1}, v_{i2}, \ldots, v_{iN})$ with
$\sum_j v_{ij} = 1$. We denote the actual welfare distribution as
$\mathbf{r} = (r_1, r_2, \ldots r_N)$. The person $i$'s welfare is
determined by the extent to which her preference is fulfilled. In other
words, we consider an equation
\begin{equation}
r_i = \mathcal{F}(\mathbf{v}_i, \mathbf{r})
\label{eq:f}
\end{equation}
with a certain function $\mathcal{F}$, which is assumed to equally apply to
all the persons. We suppose that the official wants to
announce a stable welfare distribution $\mathbf{r}$ such that each person's
relative share remains unchanged after the announcement, which means that
$\mathbf{r}$ solves Eq.~(\ref{eq:f}) self-consistently. It is important to note
that the preferences reported by each individual are assumed to be true and
available to the official at every moment, which helps us to focus on basic
ideas of the paradox.
A few remarks are in order. First, we emphasize that the official plays only a
passive role in this setup. As we will see below, the society reaches the
same self-consistent solution as long as every person's welfare becomes
public knowledge all the time. The official may guarantee such
information to be accessible and accelerate the coordination but
the official is basically assigned limited tasks compared to the original
argument.
Second, related to the first point, we do not require
the division of welfare to be impartial from a certain observer's
point of view. Our question is simply how much fulfillment one can get
depending on her preference. In this sense, our approach differs from the
impartial-division problem~\cite{clip} and does not touch conceptual
difficulties of impartiality (see, e.g., \cite{mongin}).
Finally, individuals are not behind the veil of ignorance. Rather, each of
them is supposed to construct a concrete opinion about every other individual
using any kind of available information. Although this can impose
practical difficulties in a large society,
it helps us avoid any theoretical ambiguity or conflict with the ethic of
priority~\cite{moreno} found in the veil of ignorance~\cite{veil}.

In order to give a more concrete form to Eq.~(\ref{eq:f}), we first consider how
to measure similarity or affinity between distributions and then plug it
into Eq.~(\ref{eq:f}). Suppose two arbitrary distributions, $\mathbf{p} = (p_1,
\ldots, p_N)$ and $\mathbf{q} = (q_1, \ldots, q_N)$, with $p_i \ge 0$, $q_i \ge
0$, and $\sum_i^N p_i = \sum_i^N q_i = 1$. We define a suitable affinity
function $\rho_N(\mathbf{p}, \mathbf{q})$ between them, whose specific functional
form will be characterized by requiring the following four
postulates~\cite{mathai}. First, we postulate separability, which means that
one can refine affinity contribution from a certain bin by looking into the
bin in a higher resolution without referring to the outside of the bin.
Second, we postulate invariance under permutation, because every bin is
equivalent. Third, the affinity should be non-negative, i.e.,
$\rho_N(\mathbf{p}, \mathbf{q}) \ge 0$, where $\rho_N(\mathbf{p}, \mathbf{q})=0$ if and
only if $\mathbf{p}$ is orthogonal to $\mathbf{q}$, whereas a maximum value is
obtained if and only if $\mathbf{p} = \mathbf{q}$. The distributions $\mathbf{p}$ and
$\mathbf{q}$ are orthogonal when $p_i=0$ for every non-zero $q_i$ and vice
versa. Last, it should be symmetric in the sense that $\rho_N(\mathbf{p},
\mathbf{q}) = \rho_N(\mathbf{q}, \mathbf{p})$, which is intuitively justified. These
four postulates characterize our affinity function as
\begin{equation}
\rho_N(\mathbf{p}, \mathbf{q}) \propto \sum_{i=1}^N (p_i q_i)^{1/2},
\label{eq:h}
\end{equation}
commonly known as the Bhattacharyya measure~\cite{bhatt}.
While details of the derivation are shown in Appendix,
this functional form has a clear geometric
interpretation: it can be viewed as a dot product of two vectors
$(\sqrt{x_1}, \sqrt{x_2}, \ldots \sqrt{x_N})$ and $(\sqrt{y_1}, \sqrt{y_2},
\ldots \sqrt{y_N})$, both of which are located on an $N$-dimensional unit
sphere by $\sum_i \left(\sqrt{x_i}\right)^2 = \sum_i
\left(\sqrt{y_i}\right)^2 = 1$. It therefore becomes maximized when two
vectors point in the same direction.

It is plausible to assume that the function
$\mathcal{F}(\mathbf{v}_i, \mathbf{r})$ in Eq.~(\ref{eq:f}) will be a non-decreasing
function of the affinity between $\mathbf{v}_i$ and $\mathbf{r}$, so that
$\mathcal{F}(\mathbf{v}_i, \mathbf{r}) = \mathcal{F}[ \rho_N(\mathbf{v}_i, \mathbf{r})
]$. Specifically, we infer that $\mathcal{F}(\mathbf{v}_i, \mathbf{r}) \propto
\rho_N^2(\mathbf{v}_i, \mathbf{r})$, since $\rho_N (\mathbf{v}_i, \mathbf{r})$ contains
dimensionality of $\sqrt{r_i}$ according to Eq.~(\ref{eq:h}). The precise value
of the proportionality coefficient should be determined by the normalization
condition of $\mathbf{r}$. For notational convenience, let us define $\mathbf{s}
\equiv (s_1, s_2, \ldots, s_N)$ with $s_i \equiv \sqrt{r_i}$ and $\mathbf{w}_i
\equiv (w_{i1}, w_{i2}, \ldots, w_{iN})$ with $w_{ij} \equiv \sqrt{v_{ij}}$.
Equation~(\ref{eq:f}) then leads to $s_i \propto \sum_j w_{ij} s_j$, or in a
matrix form,
\begin{equation}
\mathbf{s} = \lambda^{-1} \mathcal{W} \mathbf{s}
\label{eq:ws}
\end{equation}
where $\mathcal{W} \equiv \left\{ w_{ij} \right\}$ and $\lambda$ is for
normalizing $|\mathbf{s}|^2$, the total welfare. This formalism
is reminiscent of the quantum mechanics, where a wavefunction $\psi$ is
obtained by solving an eigenvalue problem $\mathcal{H} \psi = E \psi$ with a
Hamiltonian matrix $\mathcal{H}$ and its eigenvalue $E$. What one can
measure in experiments is probability density $|\psi|^2$.
An $N$-dimensional matrix preserving $|\mathbf{s}|^2$ is called orthogonal, and
its degrees of freedom is the number of possible planes of rotation in $N$
dimension, which is $N(N-1)/2$. Since $\mathcal{W}$ generally has $N^2$
elements and $N$ normalization conditions, it has $N(N-1)$ degrees of
freedom, so the magnitude of $\lambda$ will differ from one in general.

\section*{Results}

\subsection*{Two-person case}
The simplest example of $\mathcal{W}$ describes a situation where two
persons have not ever conceived of each other as a society member to share
welfare with. The corresponding matrix is written as
\[ \mathcal{W} = \begin{pmatrix}
1 & 0\\
0 & 1
\end{pmatrix}. \]
This identity matrix does not change the input state at all, which means
that the official cannot really coordinate these two indifferent persons'
opinions in the way that we have assumed. This is actually an example of a
reducible matrix~\cite{meyer} or a society that can be divided into smaller
pieces: $\mathcal{W}$ is irreducible if there exists a
sequence of $[k_1, k_2, \ldots, k_n]$ for any $i$ and $j$ such that
$w_{ik_1} \times w_{k_1 k_2} \times \cdots \times w_{k_n j}$ is non-zero.
Otherwise, $\mathcal{W}$ is reducible.
Such a reducible case is not our
concern since a society is meaningful only when individuals interact with
each other.
Henceforth, only irreducible cases are
considered. Then, unless everyone has zero self-interest, one can prove that
there exists a unique stable distribution $\mathbf{r}_p$ for every
$\mathcal{W}$ by using the Perron-Frobenius theorem~\cite{meyer}. In other
words, $\mathbf{r}_p$ is the only stable fixed point under the action of
$\mathcal{W}$, so the official should distribute welfare as given by
$\mathbf{r}_p$.

Let us consider a situation where an egalitarian with $\mathbf{v}_1 = (1/2,
1/2)$ meets a selfish person with $\mathbf{v}_2 = (0.01,0.99)$. The
corresponding matrix formulation will be
\begin{equation}
\mathcal{W} = \begin{pmatrix}
1/\sqrt{2} & 1/\sqrt{2} \\
1/10 & \sqrt{99}/10
\end{pmatrix}.
\label{eq:wex}
\end{equation}
and its stable welfare distribution is obtained by analyzing eigenvectors
as $\mathbf{r}_p \approx (0.72,0.28)$. We arrive at this $\mathbf{r}_p$ even if we
start from $\mathbf{r} = (0.5,0.5)$ for the following reason:
let $\mathbf{r}$ be known to both the persons every time step. The
egalitarian first feels happy to see the initial equality in $\mathbf{r}
= (0.5,0.5)$, while the selfish person feels unsatisfied, which makes a
difference at the next step. The drop in $r_2$ makes the selfish person even
more upset, so her welfare continues to decrease until it reaches the stable
value, $r_2 = 0.28$.

\begin{figure}[!ht]
\begin{center}
\includegraphics[width=0.45\textwidth]{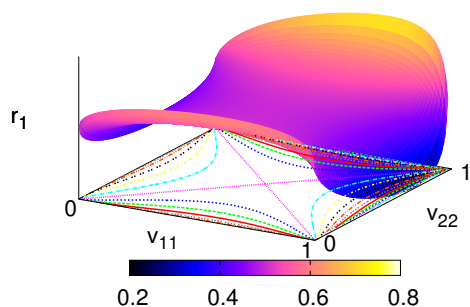}
\end{center}
\caption{{\bf Person 1's welfare in the two-person case, obtained from
Eq.~(\ref{eq:two})}. The curves on the plane show contour lines.}
\label{fig:contour}
\end{figure}

We now show that egalitarianism is the minimax solution of this two-person
zero-sum game~\cite{dav}. Generalizing Eq.~(\ref{eq:wex}) as
\begin{equation}
\mathcal{W} = \begin{pmatrix}
\sqrt{v_{11}} & \sqrt{1-v_{11}}\\
\sqrt{1-v_{22}} & \sqrt{v_{22}}
\end{pmatrix},
\label{eq:two}
\end{equation}
the eigenvalue analysis yields the converged share for the first person,
$r_1$, as shown in Fig.~\ref{fig:contour}. It is a saddle-like
shape and this person can minimize risk when she has demanded a
moderate share of $1/2$ at the first place. The same is true
for the other person as well. Although we have assumed
fixed preferences in developing the model,
if the preferences can evolve in the long
run to maximize individual welfare, therefore, this plot shows that this
two-person case will lead to an equal welfare distribution.

In practice, a selfish person can be tempted to deceive the official by
reporting a false preference to receive a larger share. Provided that person
2 has claimed her self-interest as a certain value $v_{22}$, person 1 can
always compute the best reply $\tilde{v}_{11} = b(v_{22})$ by looking up the
maximal share $\tilde{r}_1$ at the given $v_{22}$ in Fig.~\ref{fig:contour}.
Even if her true self-interest $v_{11}$ is higher than this
false $\tilde{v}_{11}$, she should still report $\tilde{v}_{11}$ to the
official, knowing that she cannot get better than $\tilde{r}_1$ in any
way. When person 1 has chosen $\tilde{v}_{11}$ for this reason, the same
consideration will lead person 2 to choose $\tilde{v}_{22} =
b(\tilde{v}_{11}) = b [b(v_{22})]$, and this reasoning can be repeated
between them {\it ad infinitum}. Such a strategic
consideration eventually forces them to choose the egalitarian preference in
common, since successive iteration of the best-reply function $b$ drives
every initial input $v_{22} \in [0,1)$ into the egalitarian fixed point,
although none of the players are really egalitarians.

\subsection*{Egalitarianism as a Nash equilibrium}
Let us consider an $N$-person case where all except one are egalitarian.
That is, $\mathbf{v}_i = (1/N, \ldots, 1/N)$ for every $i \neq 1$.
We observe that these $N-1$ persons will have exactly the same welfare since
they always get the same amount of affinity for any welfare distribution
$\mathbf{r}$. Let us thus denote every egalitarian's welfare as a single
variable $R$. Recalling the separability, we find that
all the elements $v_{1i}$ with $i \neq 1$ must be the same
in order to maximize person 1's welfare,
because her preference about the egalitarians should match with welfare
distribution among them.
Therefore, person 1 should have a preference of $\mathbf{v}_1 = (v_{11},
V, V, \cdots, V)$ with $V \equiv (1-v_{11})/(N-1)$.
As a consequence, the full $N \times N$ matrix calculation
can be simplified to the following $2 \times 2$ matrix calculation
\begin{eqnarray*}
&&\begin{pmatrix}
\sqrt{v_{11}} & (N-1)\sqrt{V}\\
\sqrt{1/N} & (N-1)\sqrt{1/N}
\end{pmatrix}
\begin{pmatrix}
\sqrt{r_1} \\ \sqrt{R}
\end{pmatrix}\\
&=&
\begin{pmatrix}
\sqrt{v_{11} r_1} + (N-1)\sqrt{VR} \\
\sqrt{r_1/N} + (N-1)\sqrt{R/N}
\end{pmatrix}
\end{eqnarray*}
with a normalization condition $r_1 + (N-1)R = 1$. The stable distribution
from the simplified matrix
then yields $r_1$ as a function of $v_{11}$, which has $\partial
r_1 / \partial v_{11} = 0$ and $r_1 = 1/N$ at $v_{11} =
1/N$. In short, the best possible preference for person 1 is an
egalitarian one. If egalitarianism is pervasive, one gets worse off by
having another type of preference, which means that egalitarianism is a
strict Nash equilibrium~\cite{nash}. This reproduces Dworkin's paradox in
mathematical terms in the sense that a non-egalitarian in an egalitarian
society will have relatively less welfare. A difference from the original
paradox is that the official cannot really compensate the non-egalitarian
within our formulation since the welfare distribution will
converge to the same point again as soon as the compensation is known
in public.

\subsection*{Inhomogeneous society}
Egalitarian preference can be still favorable even when people are all
different. For instance, people are not equally born. Let this unavoidable
inequality be described by a
uniform random variable $\zeta_i \in [-1,1]$. Person $i$'s overall political
preference can be described by another uniform random variable $\phi_i \in
(-1/N, 1/N)$: for $\phi_i = 0$, this person is an
egalitarian. If $\phi_i > 0$, she believes that the better deserve more,
while $\phi_i < 0$ means the opposite.
In addition, $\phi_i$ is assumed to be uncorrelated with $\zeta_i$.
The political preference is then assigned as
$v_{ij} = \phi_i \zeta_j + 1/N$,
which satisfies $v_{ij} > 0$.
Since $\zeta_j$ is a relative quantity, one
can always subtract an offset value to make $\sum_j \zeta_j = 0$,
by which the normalization condition $\sum_j v_{ij} = 1$ is satisfied.
We can obtain the stable distribution by taking $N$ as a very large
constant and assuming that $r_i$ is a function of $\phi_i$ only.
By replacing the summation in $\sqrt{r_i (\phi_i)} \propto
\sum_{j=1}^N \sqrt{\phi_i \zeta_j + 1/N} \sqrt{r_j (\phi_j)}$ by an
integral, we get
\[ r_i (\phi_i) = 2 \phi_i^{-2} \left[
(\phi_i + 1/N)^{3/2} - (-\phi_i + 1/N)^{3/2} \right]^2 / (3\pi + 8), \]
where the proportionality coefficient is determined by the normalization
condition. This $r_i$ is an even function of $\phi_i$ with a maximum
$r_i (\phi_i=0) = 18/(3\pi+8) \times 1/N \approx 1.033/N$,
implying the highest fulfillment for an egalitarian.
One could point out that this $r_i (\phi_i)$ describes just one possible
distribution, not necessarily the stable one. However, we see
that the integral is positive for all $\phi_i$, and thus for all $i$, and
the Perron-Frobenius theorem tells us that this positivity is true only for
the stable distribution~\cite{meyer}.
This justifies our starting assumption that person $i$'s welfare $r_i$ is not
determined by her innate part $\zeta_i$ but by her political
preference $\phi_i$ in this society.
It is notable that critics have said that the idea of
equality in welfare is insensitive to individual responsibility~\cite{dw}.
As explained in \cite{fleur}, one may consider a set of variables
characterizing an individual and classify them into two categories: the
first category consists of innate properties such as talents that an
individual is hardly responsible for. The second category, on the other
hand, includes choices and even some of preferences that we can connect to
individual responsibility.
If we regard $\zeta_i$ as representing the first category while $\phi_i$ as
representing the second category, this example shows that each individual
does take responsibility for her political preference but not for her talents.
It could be also argued that the limit $N \rightarrow \infty$ squeezes
$\phi_i \in (-1/N, 1/N)$ into zero so that the whole problem reduces to the
egalitarian society above, where everyone gets $r_i = 1/N$. That can be
regarded as a first-order approximation of this problem. The calculation
given here shows that an egalitarian indeed receives $3.3\%$ more than in
the crude approximation.

\subsection*{Homogeneously unequal preference}
\begin{figure}[!ht]
\begin{center}
\includegraphics[width=0.45\textwidth]{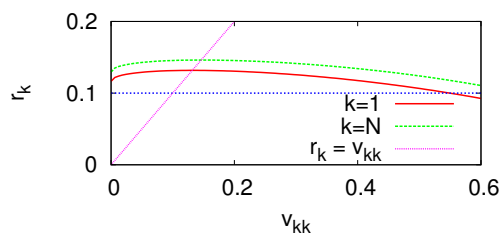}
\end{center}
\caption{{\bf Equation~(\ref{eq:rk2}) as a function of $v_{kk}$ when $h_i
\propto i^2$ for $N=10$.} The horizontal line shows $r_k = 1/N$. The maximum
of $r_k$ is located at the crossing with the line $r_k=v_{kk}$.
}
\label{fig:unequal}
\end{figure}

In all the cases considered so far, preferences could be said to
be neutral on the society level in the sense that there is no systematic
bias over the whole society.
Let us now imagine that $N-1$ persons with identical preferences,
$\mathbf{h} = (h_1, h_2, \ldots, h_N)$, but not necessarily
egalitarians. The other person indexed by $k$ has another type of
preference, $\mathbf{v}_k = (v_{k1}, v_{k2}, \ldots, v_{kN})$. By the
similar reasoning as in the egalitarian society,
the situation can be simplified to the following
$2 \times 2$ matrix:
\[ \begin{pmatrix}
\sqrt{v_{kk}} & \sqrt{(N-1)(1-v_{kk}})\\
\sqrt{h_k} & {\sum_i}'' \sqrt{h_i}
\end{pmatrix}, \]
where ${\sum_i}''$ means a summation over $i$ excluding $k$.
Recall that the $(N-1)$ persons with an identical preference have the same
amount of fulfillment so person $k$ should not distinguish them in order to
maximize affinity between her preference and the welfare distribution among
them. The above matrix means that we should only determine how to divide
welfare between the person $k$ and the other $(N-1)$ persons.
The eigenvalue analysis leads to
\begin{equation}
r_k = (1-v_{kk})(1-v_{kk} + X^2/4)
\label{eq:rk2}
\end{equation}
with
\[ X \equiv \sqrt{v_{kk}} - H - \sqrt{ (\sqrt{v_{kk}} - H)^2 + 4\sqrt{h_k
(N-1) (1-v_{kk})}} \]
and $H \equiv {\sum_i}'' \sqrt{h_i}$.
We can differentiate Eq.~(\ref{eq:rk2}) with respect to $v_{kk}$ to find the
maximum. An easier alternative way is to observe from the separability
that the maximum value is obtained when $r_k = v_{kk}$, because
the question is how to match person $k$'s preference $(v_{kk},
1-v_{kk})$ with the welfare distribution $(r_k, 1-r_k)$, where the second
elements represent the whole $(N-1)$ persons.
By solving $r_k=v_{kk}$ with Eq.~(\ref{eq:rk2}), one can get $v_{kk}$
maximizing $r_k$ as a solution of the following equation,
\begin{equation}
[h_k(N-1) + H^2] w^4 - 2H w^3 + (1-H^2) w^2 + 2H w - 1 = 0,
\label{eq:4th}
\end{equation}
with $w \equiv \sqrt{v_{kk}}$. For the egalitarian $\mathbf{h}$ with
$h_k = 1/N$ and $H = (N-1)\sqrt{1/N}$, substituting
$v_{kk} = 1/N$ satisfies Eq.~(\ref{eq:4th}), consistently with the
analysis of the $(N-1)$ egalitarians. We may also suppose that $\mathbf{h}$
describes an unequal distribution so that the whole society is biased in a
certain way. As a specific example, let us assume that $h_i \propto i^2$,
that is, almost everyone wants people with higher indices to have more
welfare. With a normalization constant, it should mean that $h_i = i^2/Z$
with $Z \equiv N(N+1)(2N+1)/6$, and we thus have
\[
H = \sum_{i=1}^N \sqrt{h_i} - \sqrt{h_k}
= \sqrt{6N(N+1)}/(2\sqrt{2N+1}) - k/\sqrt{Z}.
\]
Inserting this into Eq.~(\ref{eq:rk2}), we plot $r_k$ in
Fig.~\ref{fig:unequal}, where the maximum is found at the crossing with $r_k
= v_{kk}$. It is a little higher than $1/N$ for every $k$. We can do the
same calculation for a more severe situation of inequality by setting $h_i
\propto i^4$, which again yields the same conclusion with a bit larger
$v_{kk}$. The difference between the optimal $v_{kk}$ and
$1/N$ does not vanish as $N \rightarrow \infty$ whether $h_i
\propto i^2$ or $i^4$. This can be shown by inserting $v_{kk} =
1/N$ on the left-hand side of Eq.~(\ref{eq:4th}) and taking $N
\rightarrow \infty$, which does not yield zero on the right-hand side.
Therefore, the person $k$ should demand a little more for herself
than before and distribute the remainder of the preferences equally to the
others, even though they are far from egalitarians, in order to get the
maximum welfare.

\subsection*{Transient behavior}
\begin{figure}[!ht]
\begin{center}
\includegraphics[width=0.45\textwidth]{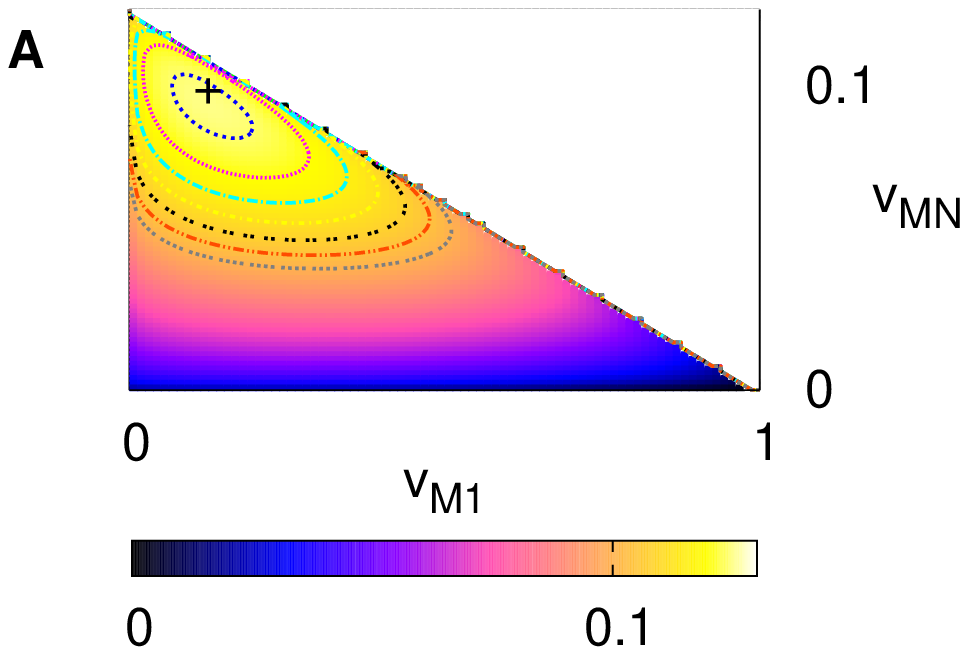}
\includegraphics[width=0.45\textwidth]{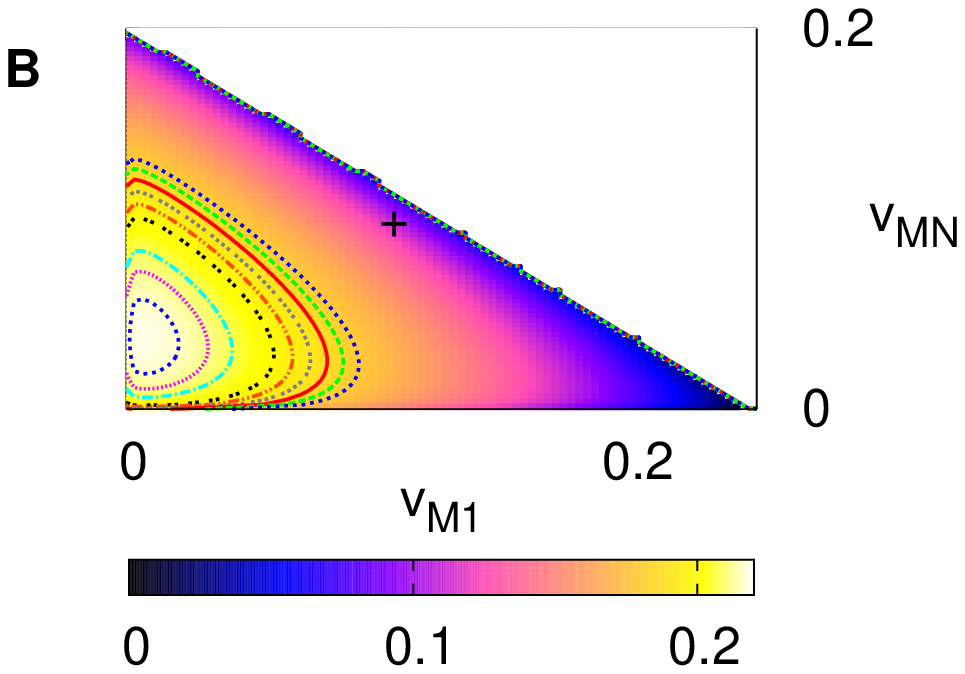}
\end{center}
\caption{{\bf 
$r_M (v_{M1}, v_{MN})$ obtained by solving Eq.~(\ref{eq:inhom}) within a
region $0 \le (M-1)v_{M1} + (N-M)v_{MN} \le 1$ for $N=10$.}
(A) $M=2$. (B) $M=5$. The crosses show $(v_{M1}, v_{MN}) = (1/N, 1/N)$.
}
\label{fig:inhom}
\end{figure}

In order to see whether egalitarians can eventually take over the
society, we need to check whether the egalitarian preference
remains as an attractive alternative when the society has both
egalitarians and non-egalitarians with significant numbers.
Let us imagine an inhomogeneous society where there are roughly two large
groups: every person in one group of size $N-M$ occupies a high index $i$
and believes that the welfare should be proportional to $i^2$. On the other
hand, every person in the other group of size $M-1$ has a low index and an
egalitarian preference. Our question is what kind of preference is good for
a person on the border, i.e., with index $i=M$. Again, since people with
identical preferences will get the same amount of welfare, the focal person
on the border need not distinguish the members in each group: suppose that
she wishes $v_{M1}$ for each member in the egalitarian group and $v_{MN}$
for each member in the non-egalitarian group. The normalization condition
then determines her self-interest $v_{MM} = 1-(M-1)v_{M1} - (N-M)v_{MN}$.
Depending on how she decides $v_{M1}$ and $v_{MN}$, her final welfare $r_M$
will be calculated by analyzing the following $3 \times 3$ matrix,
\begin{equation}
\begin{pmatrix}
(M-1)/\sqrt{N} & \sqrt{1/N} & (N-M)/\sqrt{N}\\
(M-1)\sqrt{v_{M1}} & \sqrt{v_{MM}} & (N-M)\sqrt{v_{MN}}\\
\frac{M(M-1)}{2\sqrt{Z}} & M/\sqrt{Z}
& \frac{N(N+1)}{2\sqrt{Z}} - \frac{M(M+1)}{2\sqrt{Z}}
\end{pmatrix},
\label{eq:inhom}
\end{equation}
where the second row describes this focal person $M$. The conservation of
the total welfare is imposed by setting $(M-1)r_{M1} +r_{MM}+
(N-M)r_{MN}=1$.
When $M$ is small, the maximum of $r_M$ is close to the egalitarian solution
$(v_{M1}, v_{MN}) = (1/N, 1/N)$ (Fig.~\ref{fig:inhom}A).
It agrees with the result of the homogeneously unequal preferences given
above since the egalitarian preference is still an absolute minority.
Hence, if this person $M$ can choose her own preference, the society will
possibly have one more egalitarian. As $M$ becomes larger, however, the
situation gets different in that the maximum is located far from the
egalitarian solution (Fig.~\ref{fig:inhom}B). It implies that the transition
process toward the egalitarian direction may exhibit transient
behavior, instead of being smooth all the time.

\section*{Discussion}

The theory of welfare is not an empty ideal as claimed in \cite{dw}:
dealing with a society where everyone has an identical non-egalitarian
preference, we have found that the theory recommends something very similar
to an egalitarian preference, instead of just rubber-stamping the dominant
non-egalitarian opinion. In addition, this finding shows that the egalitarian
society is in fact the \emph{only} strict Nash equilibrium. We therefore
conclude that our analysis gives a strong support to equality of welfare by
specifying which social and political conditions make it possible.

On the other hand, our conclusion implies that a society can encourage
egalitarianism by guaranteeing freedom of communication so that everyone can
constantly express her fulfillment in public. In this respect, we can
perhaps mention one of the central messages in \cite{dw} that ``liberty is
essential to any process in which equality is defined and secured.'' In
particular, we would like to put an extra emphasis on the communicative
aspect of the liberty.

On a longer perspective, our results suggest an explanation of how
the concept of fairness could develop at a certain moment in the history of
evolution when human beings became able to construct internal expectation
for the future and understand others' minds by communication.
It is also worth stressing that
our conclusion on egalitarianism as a strict Nash equilibrium under
certain well-defined conditions is strong enough to open further theoretical
extensions and empirical tests.

\section*{Acknowledgments}
We are grateful to an anonymous reviewer for many helpful comments.
S.K.B. is indebted to Petter Minnhagen and Sang Hoon Lee for discussions.
S.K.B. acknowledges the support from the Swedish Research Council with
Grant No. 621-2008-4449.
J.K.C. was supported by Korea Research Foundation Grant funded by the
Korean Government (NRF-2010-330-B00077).
B.J.K. was supported by the National Research
Foundation of Korea (NRF) grant funded by the Korea government (MEST) (No.
2010-0008758).


\section*{Figure Legends}

\clearpage
\section*{Appendix: derivation of the affinity function}

Let us consider two arbitrary distributions, $\boldsymbol{p} = (p_1,
\ldots, p_N)$ and $\boldsymbol{q} = (q_1, \ldots, q_N)$, with $p_i \ge 0$,
$q_i \ge 0$, and $\sum_i^N p_i = \sum_i^N q_i = 1$. We will define
an affinity function $\rho_N(\boldsymbol{p}, \boldsymbol{q})$
between them, whose specific functional form will be characterized by
requiring the following postulates.

\begin{itemize}
\item P1: Separability
\begin{eqnarray*} 
&&\rho_N \begin{pmatrix} p_1, \ldots, p_N\\ q_1, \ldots, q_N \end{pmatrix}
= \rho_{N-k+1} \begin{pmatrix} P_k, p_{k+1}, \ldots, p_N\\ Q_k, q_{k+1},
\ldots, q_N \end{pmatrix}\\
&&+ \Delta(P_k, Q_k) \left[
\rho_k \begin{pmatrix} \frac{p_1}{P_k}, \ldots, \frac{p_k}{P_k} \\
\frac{q_1}{Q_k}, \ldots, \frac{q_k}{Q_k} \end{pmatrix} - 1
\right],
\end{eqnarray*}
where $P_k \equiv p_1 + p_2 + \cdots + p_k$ and $Q_k \equiv
q_1 + q_2 + \cdots + q_k$ are positive with $1<k<N$. In addition,
$\Delta(P_k,Q_k)$ is a non-negative differentiable function and converges to
zero as $P_k \rightarrow 0$ or $Q_k \rightarrow 0$.
The situation can be described as follows:
we first observe $\mathbf{p}' = (P_k, p_{k+1}, \ldots, p_N)$
and $\mathbf{q}' = (Q_k, q_{k+1}, \ldots, q_N)$, instead of the real
$\mathbf{p}$ and $\mathbf{q}$. We can calculate affinity
$\rho_{N-k+1}$ in this resolution.
The question is how affinity will change when we come to know the real
$\mathbf{p}$ and $\mathbf{q}$.
If the substructures inside $P_k$ and $Q_k$ have an exact affinity
by $\rho_k = 1$, nothing will change from the previous calculation,
i.e., $\rho_N = \rho_{N-k+1}$. If they actually had no affinity by $\rho_k =
0$ in this better resolution, on the other hand, the overall affinity should
decrease by a certain amount $\Delta$, which will be a function of
$P_k$ and $Q_k$. For example, for $P_k \ll 1$
or $Q_k \ll 1$, the decrement in $\rho_N$ will be also vanishingly small
even if the subpopulations inside them look completely different.
\item
P2: Invariance under permutation
\[
\rho_3 \begin{pmatrix} p_1, p_2, p_3 \\ q_1, q_2, q_3 \end{pmatrix}
= \rho_3 \begin{pmatrix} p_2, p_1, p_3 \\ q_2, q_1, q_3 \end{pmatrix}
= \ldots
= \rho_3 \begin{pmatrix} p_3, p_2, p_1 \\ q_3, q_2, q_1 \end{pmatrix}.
\]
Note that it is for $N=3$, which is enough to prove the same property
for general $N$ in combination with the other postulates.
\item
P3: Non-negativity
\[ \rho_N(\boldsymbol{p}, \boldsymbol{q}) \ge 0, \]
where $\rho_N(\boldsymbol{p}, \boldsymbol{q})=0$ if
and only if $\boldsymbol{p}$ is orthogonal to $\boldsymbol{q}$, whereas
a maximum value is obtained if and only if $\boldsymbol{p} =
\boldsymbol{q}$.
\item
P4: Symmetry
\[ \rho_N(\boldsymbol{p}, \boldsymbol{q}) =
\rho_N(\boldsymbol{q}, \boldsymbol{p}). \]
\end{itemize}
Let us explain some direct consequences of P1. For
$N=2$, it yields a trivial equality. For $N=3$, we have
\[
\rho_3 \begin{pmatrix} p_1, p_2, p_3\\ q_1, q_2, q_3 \end{pmatrix} =
\rho_2 \begin{pmatrix} P_2, p_3\\ Q_2, q_3 \end{pmatrix}
+ \Delta(P_2, Q_2) \left[
\rho_2 \begin{pmatrix} \frac{p_1}{P_2}, \frac{p_2}{P_2}\\ \frac{q_1}{Q_2},
\frac{q_2}{Q_2} \end{pmatrix} - 1
\right].
\]
For $N=4$, we will see two different ways to calculate affinity between
$(P_3,p_4)$ and $(Q_3,q_4)$. The first way is the following:
\begin{eqnarray}
&&\rho_4 \begin{pmatrix} p_1, \ldots, p_4\\ q_1, \ldots, q_4
\end{pmatrix}\nonumber\\
&=& \rho_2 \begin{pmatrix} P_3, p_4\\ Q_3, q_4 \end{pmatrix}
+ \Delta(P_3,Q_3)
\left[ \rho_3 \begin{pmatrix} \frac{p_1}{P_3}, \frac{p_2}{P_3},
\frac{p_3}{P_3} \\ \frac{q_1}{Q_3}, \frac{q_2}{Q_3}, \frac{q_3}{Q_3}
\end{pmatrix}  - 1 \right] \nonumber\\
&=& \rho_2 \begin{pmatrix} P_3, p_4\\ Q_3, q_4 \end{pmatrix}
+ \Delta(P_3,Q_3) \left\{
\rho_2 \begin{pmatrix} \frac{P_2}{P_3}, \frac{p_3}{P_3}\\
\frac{Q_2}{Q_3}, \frac{q_3}{Q_3} \end{pmatrix} \right. \nonumber\\
&&\left. + \Delta\left( \frac{P_2}{P_3}, \frac{Q_2}{Q_3} \right)
\left[
\rho_2 \begin{pmatrix} \frac{p_1}{P_2}, \frac{p_2}{P_2}\\ \frac{q_1}{Q_2},
\frac{q_2}{Q_2} \end{pmatrix} - 1
\right] - 1
\right\}. \label{eq:rho4a}
\end{eqnarray}
We then consider another way to calculate the same quantity as follows:
\begin{eqnarray}
&&\rho_4 \begin{pmatrix} p_1, \ldots, p_4\\ q_1, \ldots, q_4
\end{pmatrix}\nonumber\\ &=&
\rho_3 \begin{pmatrix} P_2, p_3, p_4\\ Q_2, q_3, q_4 \end{pmatrix}
+ \Delta(P_2,Q_2)
\left[ \rho_2 \begin{pmatrix} \frac{p_1}{P_2}, \frac{p_2}{P_2} \\
\frac{q_1}{Q_2}, \frac{q_2}{Q_2} \end{pmatrix}  - 1 \right] \nonumber\\
&=& \rho_2 \begin{pmatrix} P_3, p_4\\ Q_3, q_4 \end{pmatrix}
+ \Delta(P_3, Q_3)
\left[ \rho_2 \begin{pmatrix} \frac{P_2}{P_3}, \frac{p_3}{P_3}\\
\frac{Q_2}{Q_3}, \frac{q_3}{Q_3} \end{pmatrix} -1 \right]\nonumber\\
&&+ \Delta(P_2, Q_2)
\left[ \rho_2 \begin{pmatrix} \frac{p_1}{P_2}, \frac{p_2}{P_2} \\
\frac{q_1}{Q_2}, \frac{q_2}{Q_2} \end{pmatrix}  - 1 \right].
\label{eq:rho4b}
\end{eqnarray}
Comparing Eq.~(\ref{eq:rho4a}) with Eq.~(\ref{eq:rho4b}), we see that
\begin{equation}
\Delta(P_3, Q_3) \Delta \left(\frac{P_2}{P_3}, \frac{Q_2}{Q_3} \right)
= \Delta(P_2, Q_2).
\label{eq:delta}
\end{equation}
Note that $P_2/P_3$ can take any value between zero and one, regardless of
$P_3$, and that $Q_2/Q_3$ is also independent of $Q_3$ in the same way, as
long as $P_3$ and $Q_3$ are nonzero.

Now we define
\begin{equation}
g(x,y) \equiv \rho_2 \begin{pmatrix}x, 1-x \\ y, 1-y \end{pmatrix} - 1,
\label{def:g}
\end{equation}
for $x, y \in [0,1]$.
From P1 and P2, it is straightforward to verify
\begin{equation}
g(x,y) = g(1-x, 1-y),
\label{eq:g1x}
\end{equation}
since
\begin{eqnarray*}
&&\rho_3 \begin{pmatrix} p_1, p_2, p_3 \\ q_1, q_2, q_3 \end{pmatrix}
= \rho_3 \begin{pmatrix} p_2, p_1, p_3 \\ q_2, q_1, q_3 \end{pmatrix}\\
&=& \rho_2 \begin{pmatrix} P_2, p_3 \\ Q_2, q_3 \end{pmatrix}
+ \Delta(P_2, Q_2) \left[ \rho_2 \begin{pmatrix} \frac{p_1}{P_2},
\frac{p_2}{P_2} \\ \frac{q_1}{Q_2}, \frac{q_2}{Q_2} \end{pmatrix} -1
\right]\\
&=& \rho_2 \begin{pmatrix} P_2, p_3 \\ Q_2, q_3 \end{pmatrix}
+ \Delta(P_2, Q_2) \left[ \rho_2 \begin{pmatrix} \frac{p_2}{P_2},
\frac{p_1}{P_2} \\ \frac{q_2}{Q_2}, \frac{q_1}{Q_2} \end{pmatrix} -1
\right].
\end{eqnarray*}
Applying P1 to another permutation
\[
\rho_3 \begin{pmatrix} p_1, p_2, p_3 \\ q_1, q_2, q_3 \end{pmatrix}
= \rho_3 \begin{pmatrix} p_3, p_2, p_1 \\ q_3, q_2, q_1 \end{pmatrix},
\]
we are directly led to
\begin{eqnarray*}
&&g(p_3, q_3) + \Delta(1-p_3, 1-q_3) g\left( \frac{p_1}{1-p_3},
\frac{q_1}{1-q_3} \right)\\
&=& g(p_1, q_1) + \Delta(1-p_1, 1-q_1) g\left( \frac{p_3}{1-p_1},
\frac{q_3}{1-q_1} \right),
\end{eqnarray*}
where $p_1, p_3, q_1, q_3 \in [0,1)$.
Let us redefine the variables as $\frac{p_1}{1-p_3} \equiv p$,
$\frac{q_1}{1-q_3} \equiv q$, $1-p_3 \equiv r$, and $1-q_3 \equiv s$. Each
of these new variables $p$, $q$, $r$, and $s$ can take an arbitrary value
independently of one another, as long as $p,q \in [0,1]$ and $r,s \in
(0,1)$. The above equality can be then rewritten as
\begin{eqnarray}
&g(r,s) + \Delta(r,s) g(p,q)\nonumber\\
&=g(pr,qs) + \Delta(1-pr,1-qs)  g\left( \frac{1-r}{1-pr}, \frac{1-s}{1-qs}
\right).
\label{eq:grs}
\end{eqnarray}
Defining a function
\begin{equation}
f(p,q,r,s) \equiv g(r,s) + \left[ \Delta(r,s) + \Delta(1-r,1-s) \right] g(p,q)
\label{def:f}
\end{equation}
for $p, q, r, s \in (0,1)$,
we apply Eq.~(\ref{eq:grs}) to $f(p,q,r,s)$ to obtain
\begin{eqnarray*}
f(p,q,r,s) &=&
g(pr,qs)\\
&&+ \Delta(1-pr,1-qs)
g\left( \frac{1-r}{1-pr}, \frac{1-s}{1-qs} \right)\\
&&+ \Delta(1-r,1-s) g(p,q).
\end{eqnarray*}
Note from Eq.~(\ref{eq:delta}) that
\begin{eqnarray*}
&& \Delta(p_2+p_3, q_2+q_3) \Delta \left(\frac{p_3}{p_2+p_3},
\frac{q_3}{q_2+q_3} \right)\\
&=& \Delta(1-pr, 1-qs) \Delta \left( \frac{1-r}{1-pr}, \frac{1-s}{1-qs}
\right)\\
&=& \Delta(1-r, 1-s).
\end{eqnarray*}
Hence, we see that
\begin{eqnarray}
&&f(p,q,r,s) \nonumber\\
&=& g(pr,qs) + \Delta(1-pr, 1-qs)
g\left( \frac{1-r}{1-pr}, \frac{1-s}{1-qs} \right)\nonumber\\
&&+ \Delta(1-pr, 1-qs) \Delta \left( \frac{1-r}{1-pr}, \frac{1-s}{1-qs}
\right) g(p,q) \nonumber\\
&=& g(pr,qs)
+ \Delta(1-pr, 1-qs) \left[ g\left( \frac{1-r}{1-pr}, \frac{1-s}{1-qs}
\right) \right.\nonumber\\
&& \left. + \Delta \left( \frac{1-r}{1-pr}, \frac{1-s}{1-qs} \right) g(p,q)
\right] \label{eq:f1}\\
&=& g(pr,qs)
+ \Delta(1-pr, 1-qs) \left\{ g\left[ \frac{p(1-r)}{1-pr},
\frac{q(1-s)}{1-qs} \right] \right.\nonumber\\
&&+ \left. \Delta \left( \frac{1-p}{1-qr}, \frac{1-q}{1-qs} \right) g(r,s)
\right\} \label{eq:f2},
\end{eqnarray}
where the last equality comes from Eq.~(\ref{eq:grs}).
Equation~(\ref{eq:g1x}) then gives us
\[
g\left[ \frac{p(1-r)}{1-pr}, \frac{q(1-s)}{1-qs} \right]
= g\left( \frac{1-p}{1-pr}, \frac{1-q}{1-qs} \right),
\]
relating Eqs.~(\ref{eq:f1}) and (\ref{eq:f2}) as follows:
\begin{eqnarray*}
&&f(p,q,r,s)\\
&=& g(pr,qs)
+ \Delta(1-pr, 1-qs) \left[ g\left( \frac{1-r}{1-pr}, \frac{1-s}{1-qs}
\right) \right.\\
&& + \left. \Delta \left( \frac{1-r}{1-pr}, \frac{1-s}{1-qs} \right) g(p,q)
\right]\\
&=& g(pr,qs)
+ \Delta(1-pr, 1-qs) \left[ g\left( \frac{1-p}{1-pr}, \frac{1-q}{1-qs}
\right) \right.\\
&&+ \left. \Delta \left( \frac{1-p}{1-pr}, \frac{1-q}{1-qs} \right) g(r,s)
\right].
\end{eqnarray*}
In short, we have just confirmed that
\[f(p,q,r,s) = f(r,s,p,q). \]
From the definition of $f(p,q,r,s)$ in Eq.~(\ref{def:f}), we see that
\[
\frac{g(r,s)}{g(p,q)} = \frac{\Delta(r,s) + \Delta(1-r,1-s) - 1}{\Delta(p,q)
+ \Delta(1-p,1-q) - 1}.
\]
For this to be true for every independent set of $(p,q,r,s)$, it must be
that
\[ g(r,s) = C [\Delta(r,s) + \Delta(1-r,1-s) - 1] \]
with a certain constant $C$. Then P1 says that
\[ \lim_{ \substack{r \rightarrow 0\\s \rightarrow 1}} g(r,s) = -C,\]
which is the lowest possible value of $g(r,s)$.
However, according to the definition of $g(r,s)$ in Eq.~(\ref{def:g}) and P3,
this lower bound should be $-1$. In other words, $C=1$ and
\[ \rho_2 \begin{pmatrix} r, 1-r \\ s, 1-s\end{pmatrix}
= \Delta(r,s) + \Delta(1-r,1-s). \]
We have restricted our variables as $r \in (0,1)$ and $s \in
(0,1)$, but it is not difficult to check that the above expression can be
readily used in $r \in [0,1]$ and $s \in [0,1]$ when $\Delta$ is
characterized as will be discussed below.
Then P1 works as a recursive relation, yielding an
expression for general $N$:
\begin{equation}
\rho_N = \sum_{i=1}^N \Delta(p_i, q_i).
\label{eq:rho}
\end{equation}
Now we ask ourselves how $\Delta$ should look. P1 and P4
imply that $\Delta$ can be expanded as
$\Delta(p_i,q_i) = c_1 (p_i q_i)^{\alpha_1} + c_2 (p_i q_i)^{\alpha_2} +
\cdots$ with coefficients $c_n$ and exponents $\alpha_n > 0$. So let us write
\[ \rho_N = \sum_{i=1}^N \sum_{n=1}^{\infty} c_n (p_i q_i)^{\alpha_n}. \]
If we define
$l_{\alpha}(\boldsymbol{p}, \boldsymbol{q}) \equiv \sum_{i=1}^N (p_i
q_i)^{\alpha}$, it means that $\rho_N$ should be of the following form,
\begin{equation}
\rho_N (\boldsymbol{p}, \boldsymbol{q}) = \sum_{n=1}^{\infty}
c_n l_{\alpha_n} (\boldsymbol{p}, \boldsymbol{q}).
\label{eq:over}
\end{equation}
Among every possible $l_{\alpha}$, it is only $l_{1/2}$ that satisfies
the maximization condition in P3. This can be shown by
variational calculus using a Lagrange multiplier $\mu$,
\[
\frac{\partial}{\partial q_i}
\left[ l_{\alpha}(\boldsymbol{p}, \boldsymbol{q}) -
\mu \sum_{j=1}^N q_j \right]_{q_i=p_i} = \alpha p_i^{2\alpha-1} - \mu = 0,
\]
which is satisfied for any $p_i$ when $\alpha=1/2$ and
$\mu=\alpha$. Hence we separate this term from the others in
Eq.~(\ref{eq:over}) as
\[
\rho_N (\boldsymbol{p}, \boldsymbol{q})
= c l_{1/2}(\boldsymbol{p}, \boldsymbol{q})
+ {\sum_n}' c_n l_{\alpha_n} (\boldsymbol{p}, \boldsymbol{q}),
\]
where $\sum'$ means that $\alpha_n=1/2$ is excluded from the summation.
There exists a maximum value in $\rho_N$, which is obtained by
\begin{eqnarray*}
\rho_N (\boldsymbol{p}, \boldsymbol{p}) &=& c l_{1/2}(\boldsymbol{p},
\boldsymbol{p}) + {\sum_n}' c_n l_{\alpha_n} (\boldsymbol{p}, \boldsymbol{p})\\
&=& c + {\sum_n}' c_n \sum_i p_i^{2 \alpha_n},
\end{eqnarray*}
noting that $l_{1/2}(\boldsymbol{p}, \boldsymbol{p}) = \sum_i p_i = 1$.
Since this value is the same for every $\boldsymbol{p}$, the last
term should be kept constant for any $p_i$, and we have to choose
$c_n = 0$ for every $n$ in the second term.
To sum up, our affinity function is characterized as
\[ \rho_N(\boldsymbol{p}, \boldsymbol{q}) = c \sum_{i=1}^N (p_i q_i)^{1/2} \]
with a positive constant $c$, the maximum value of $\rho_N$.

\end{document}